# Compound Multiple Access Channel with Common Message and Intersymbol Interference

Mostafa Monemizadeh, Saeed Hajizadeh, Seyed Alireza Seyedin, and Ghosheh Abed Hodtani
Department of Electrical Engineering
Ferdowsi University of Mashhad
Mashhad, Iran
mostafamonemizadeh@gmail.com; seyedin@um.ac.ir; ghodtani@gmail.com; saeed.hajizadeh1367@gmail.com

*Abstract*—In this paper, we characterize the capacity region for the two-user linear Gaussian compound Multiple Access Channel with common message (MACC) and with intersymbol interference (ISI) under an input power constraint. The region is obtained by converting the channel to its equivalent memoryless one by defining an *n*-block memoryless circular Gaussian compound MACC model and applying the discrete Fourier transform (DFT) to decompose the *n*-block channel into a set of independent parallel channels whose capacities can be found easily. Indeed, the capacity region of the original Gaussian compound MACC equals that of the *n*-block circular Gaussian compound MACC in the limit of infinite block length. Then by using the obtained capacity region, we derive the capacity region of the strong interference channel with common message and ISI.

## I. INTRODUCTION

In digital communication systems, intersymbol interference (ISI) is an unavoidable and undesirable property and a main source of performance degradation in which symbols interfere with the following or preceding transmitted symbols. The major causes of the ISI are multipath fading and the transmission of a signal through a bandlimited channel. Since the channel with ISI is a channel with memory, the capacity characterization of such channel is not easy. The capacity of a single-user Gaussian channel with ISI was derived in [1], where the idea of converting such channel to its equivalent memoryless one by using *n*-block memoryless circular Gaussian channel model and applying the discrete Fourier transform (DFT) to decompose the *n*-block channels into independent channels whose capacities can be found easily, was introduced. Up to now, this idea is the basis of all approaches to obtain the capacity regions of the various channels with ISI. The capacity region of a two-user linear Gaussian multiple access channel (MAC) with ISI was found in [2] using the same methodology in [1]. Goldsmith and Effros [3] obtained the capacity region of a finite-memory broadcast channel (BC) with ISI and colored Gaussian noise and showed that this capacity region is equal to the capacity region of an *n*-circular Gaussian BC as *n* grows to infinity. Recently, Choudhuri and Mitra [4] derived single-letter expressions for the achievable rates and an upper bound on the capacity of a relay channel with ISI and additive colored Gaussian noise.

In this paper we derive the capacity region of the two-user Gaussian compound MAC with common message (GCMACC) and with ISI. The two-user compound MAC is a communication channel with two senders and two receivers in which two senders communicate with two receivers simultaneously and each of the two receivers need to decode all messages sent from both senders [6]–[9]. The capacity regions of the compound MAC with a common message and of compound MAC with conferencing encoders were established in [6]. Simeone *et al.* [7] extended the channel models in [6] to the case with conferencing decoders and studied the compound MAC with a common message and conferencing decoders, and compound MAC with both conferencing encoders and decoders. Recently, the compound MACC with Slepian-Wolf type correlated channel states has been studied in [8].

We obtain the capacity region of the Gaussian compound MACC with ISI using the same approach that has been used to obtain the capacity of the single-user and synchronous multi-user channels with ISI [1]–[5]. We first define a similar channel model, *n*-block memoryless circular Gaussian compound MACC that can be decomposed into a set of *n*-parallel, memoryless and independent scalar channels (in DFT domain) whose individual capacities are given by prior results. Then we obtain the capacity of this *n*-block memoryless circular Gaussian model based on the DFT decomposition. Finally, the capacity region of the original Gaussian compound MACC is derived using this fact that this capacity region equals that of the *n*-block circular Gaussian compound MACC in the limit of infinite block length [1]–[5]. Since both senders are allowed to transmit a common message cooperatively, the channel is block (or frame)-synchronous and we can employ this fact that any synchronous multiterminal channel and its circular approximation have the same capacity region [1]–[5]. One of the benefits of study the Gaussian compound MACC with ISI is that we can easily derive the capacity region of the strong interference channel with common message (SICC) and with ISI as a special case of it. The rest of the paper is organized as follows. In Section II, we define the linear GCMACC with ISI and *n*-block circular GCMACC. All the main results are presented in Section III. In Section IV, we prove the derived capacity region in Theorem 1.

## II. DEFINITIONS AND CHANNEL MODEL

We use notations and formulations similar to [3] and [4]. We use $*$ and $\circledast$ to denote the linear and circular convolution, respectively. $\langle A \rangle_n$ equals $A$ modulo $n$ except when $A$ is zero or an integer multiple of $n$, in which case $\langle A \rangle_n = n$. We denote sequence $(\ldots, s_1, s_2, \ldots)$ by $\{s\}$, subsequence $(s_a, \ldots, s_b)$ as $\{s_k\}_{k=a}^{b}$ and vector $(s_1, \ldots, s_n)$ by $s^n$. $(\cdot)^\dagger$ denotes the conjugate transpose of $(\cdot)$ and $\bar{\alpha} = 1 - \alpha$. Also for a matrix $\boldsymbol{A}$, $|\boldsymbol{A}|$ denotes the absolute value of the determinant of $\boldsymbol{A}$.

We consider the discrete-time linear Gaussian compound MACC with ISI shown in Fig. 1, where there are different four sets of ISI coefficients $\{h_{pq,t}\}_{t=0}^{m}, p,q = 1,2$, one for each link. In this channel, the output sequences $\{y_{1,k}\}$ and $\{y_{2,k}\}$ are ($-\infty < k < \infty$)

$$y_{1,k} = \sum_{t=0}^{m}(h_{11,t}x_{1,k-t} + h_{21,t}x_{2,k-t}) + z_{1,k}$$
$$= h_{11,k} * x_{1,k} + h_{21,k} * x_{2,k} + z_{1,k} \quad (1)$$

$$y_{2,k} = \sum_{t=0}^{m}(h_{12,t}x_{1,k-t} + h_{22,t}x_{2,k-t}) + z_{2,k}$$
$$= h_{12,k} * x_{1,k} + h_{22,k} * x_{2,k} + z_{2,k} \quad (2)$$

where $\{x_{1,k}\}$ and $\{x_{2,k}\}$ are input sequences transmitted by user 1 and user 2, respectively, and $\{z_{1,k}\}$ and $\{z_{2,k}\}$ are zero-mean stationary Gaussian noise processes with autocorrelation functions $R_1[t]$ and $R_2[t]$, respectively. These autocorrelation functions are assumed to have a common finite support $t_{max}$, i.e., $R_1[t] = R_2[t] = 0$ for $|t| \geq t_{max}$. Also, we assume that all channel impulse responses $\{h_{pq,k}\}_{k=0}^{m}, p,q = 1,2$; have common memory length $m$. We only consider the case $m \geq t_{max}$. For the case $m < t_{max}$, the channel impulse responses can be zero padded to make them equal. Since the outputs are linear combinations of the inputs, for a given $m$, this channel is called the linear Gaussian compound MACC (LGCMACC) with finite memory $m$. Moreover, since the channel outputs at a time instance depend on the input symbols of that time as well as previous input symbols, the channels have ISI.

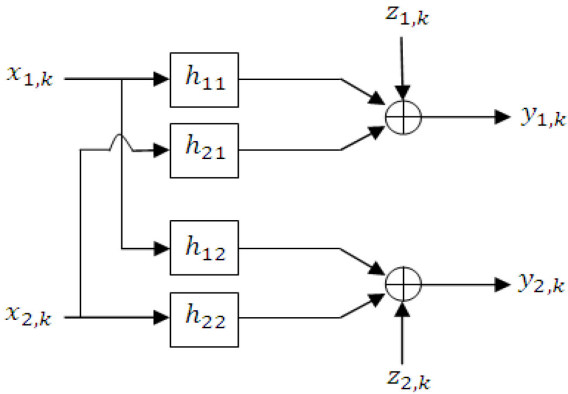

Figure 1. The discrete-time linear Gaussian CMACC with ISI.

The input sequences are subjected to the following average power constraints for all $n$:

$$\frac{1}{n}\sum_{k=1}^{n}E[x_{q,k}^2] \leq P_q, \quad q = 1,2. \quad (3)$$

The transfer functions of the channel links are (the DFTs of channel impulse responses)

$$H_{pq} = H_{pq}(\omega) = \sum_{t=0}^{m}h_{pq,t}e^{-j\omega t}, \quad p,q \in \{1,2\} \quad (4)$$

which are periodic in $\omega$ with period $2\pi$. The spectral noise densities of the channels are

$$N_q = N_q(\omega) = \sum_{t=-(m-1)}^{m-1}R_q[t]e^{-j\omega t}, \quad p,q \in \{1,2\} \quad (5)$$

To compute the capacity region of the LGCMACC with ISI, we modify the LGCMACC (with memory $m$) and define an equivalent *n-block memoryless circular* channel model called *n-block circular Gaussian CMACC (n-CGCMACC)*, for $n > m$, in which the channel outputs over any *n*-block transmission are independent of channel inputs and noise samples of other *n*-blocks. Specifically, the *n*-CGCMACC over each *n*-block has input vectors $\{x_{pq,k}\}_{k=1}^{n}, p,q \in \{1,2\}$ which produce output vectors $\{\tilde{y}_{1,k}\}_{k=1}^{n}$ and $\{\tilde{y}_{2,k}\}_{k=1}^{n}$ at the first and second receivers, respectively, so that for $1 \leq k \leq n$

$$\tilde{y}_{1,k} = \sum_{t=0}^{n-1}(\tilde{h}_{11,t}x_{1,\langle k-t\rangle_n} + \tilde{h}_{21,t}x_{2,\langle k-t\rangle_n}) + \tilde{z}_{1,k}$$
$$= \tilde{h}_{11,k} \circledast x_{1,k} + \tilde{h}_{21,k} \circledast x_{2,k} + \tilde{z}_{1,k} \quad (6)$$

$$\tilde{y}_{2,k} = \sum_{t=0}^{n-1}(\tilde{h}_{12,t}x_{1,\langle k-t\rangle_n} + \tilde{h}_{22,t}x_{2,\langle k-t\rangle_n}) + \tilde{z}_{2,k}$$
$$= \tilde{h}_{12,k} \circledast x_{1,k} + \tilde{h}_{22,k} \circledast x_{2,k} + \tilde{z}_{2,k} \quad (7)$$

where, $\{\tilde{h}_{pq,k}\}_{k=0}^{n-1} = (h_{pq,0}, h_{pq,1}, \ldots, h_{pq,m}, 0, \ldots, 0)$ for $p,q \in \{1,2\}$, i.e., $\{\tilde{h}_{pq,k}\}_{k=0}^{n-1}$ is an extended version of $\{h_{pq,k}\}_{k=0}^{m}$ which is extended with $(n - m - 1)$ zeros. Note that in (6) and (7), the channel impulse responses are fixed vectors and input vectors are circular. We can obtain similar results by considering the fixed input vectors and circular channel impulse response vectors. Since the *n*-CGCMACC is an *n*-block memoryless channel, the noise samples $\tilde{z}_{1,k}$ and $\tilde{z}_{2,k}$ are *n*-block independent with the same means and variances as $z_{1,k}$ and $z_{2,k}$, respectively, which their autocorrelation functions $\tilde{R}_1[t]$ and $\tilde{R}_2[t]$ are periodic repetitions of $R_1[t]$ and $R_2[t]$, respectively, for noise samples within an *n*-block. The same average power constraints (3) are assumed for *n*-CGCMAC.

Now, using the DFT, we decompose the *n*-CGCMACC with ISI into a set of *n*-parallel, memoryless and independent scalar GCMACC channels in DFT domain which the capacities of these independent channels can be found easily by prior results. Therefore, applying the DFT to both sides of (6) and (7) we obtain

$$\tilde{Y}_{1,k} = \tilde{H}_{11,k}\tilde{X}_{1,k} + \tilde{H}_{21,k}\tilde{X}_{2,k} + \tilde{Z}_{1,k} \quad (8)$$
$$\tilde{Y}_{2,k} = \tilde{H}_{12,k}\tilde{X}_{1,k} + \tilde{H}_{22,k}\tilde{X}_{2,k} + \tilde{Z}_{2,k} \quad (9)$$

where for $1 \leq k \leq n$ and $p,q \in \{1,2\}$, $\tilde{Y}_{q,k}$, $\tilde{H}_{pq,k}$, $\tilde{X}_{q,k}$, and $\tilde{Z}_{q,k}$ are the DFTs of $\tilde{y}_{q,k}$, $\tilde{h}_{pq,k}$, $\tilde{x}_{q,k}$, and $\tilde{z}_{q,k}$, respectively. Consequently, the $n$-CGCMACC is equivalent to a set of $n$ parallel compound MACCs which the $k$th-component channel is as shown in Fig. 2.

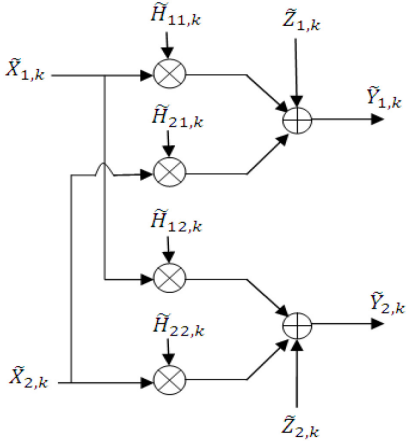

Figure 2. The $k$th-component channel.

## III. MAIN RESULTS

In this section, we first obtain the capacity region of the $n$-CGCMACC which is the same as the capacity region of the LGCMACC with ISI in the limit of infinite block length. Then by using the obtained capacity region, we derive the capacity region of the linear Gaussian strong interference channel with common message (SICC) and with ISI.

### A. Capacity Region of the Linear Gaussian Compound MAC with Common Message and ISI

Let $\mathcal{C}$ and $\mathcal{C}_n$ denote the capacity regions of the LGCMACC with finite memory and the $n$-CGCMACC, respectively. Since both senders are allowed to transmit a common message cooperatively, the channel is block (or frame)-synchronous. As our channel is a special case of a synchronous multi-terminal channel, we can apply the results in [3] and [5] and obtain the capacity region of the LGCMACC with ISI which is the same as the capacity region of the $n$-CGCMACC in the limit as $n$ goes to infinity. Note that considering the time-sharing principle [5], the capacity region of the block-synchronous compound MACC with finite memory is a convex set. Applying the results in [3] and [5] we have:

$$\mathcal{C} = \text{closure}\left(\liminf_{n\to\infty}\mathcal{C}_n\right) = \text{closure}\left(\limsup_{n\to\infty}\mathcal{C}_n\right) \quad (10)$$

where $\mathcal{C}_n = \mathcal{C}_n(P_1,P_2)$ is

$$\bigcup \left\{ \begin{array}{l} (R_0, R_1, R_2): R_0 \geq 0, R_1 \geq 0, R_2 \geq 0, \\ R_1 \leq \min\frac{1}{n}\{I(x_1^n; y_1^n | x_2^n, u^n), I(x_1^n; y_2^n | x_2^n, u^n)\} \\ R_2 \leq \min\frac{1}{n}\{I(x_2^n; y_1^n | x_1^n, u^n), I(x_2^n; y_2^n | x_1^n, u^n)\} \\ R_1 + R_2 \leq \min\frac{1}{n}\{I(x_1^n, x_2^n; y_1^n | u^n), I(x_1^n, x_2^n; y_2^n | u^n)\} \\ R_0 + R_1 + R_2 \leq \min\frac{1}{n}\{I(x_1^n, x_2^n; y_1^n), I(x_1^n, x_2^n; y_2^n)\} \end{array} \right\}$$
(11)

where the union is over all input vectors $x_1^n$ and $x_2^n$ subjected to the average power constraints

$$\text{tr}\left(E\left[x_q^n(x_q^n)^\dagger\right]\right) \leq nP_q, \quad q = 1,2. \quad (12)$$

Indeed since the $n$-CGCMACC defined in (6) and (7) is an $n$-block memoryless compound MACC, its capacity region (11) follows directly from [6] provided that we replace $(X_1, X_2, U, Y_1, Y_2)$ by $(x_1^n, x_2^n, u^n, y_1^n, y_2^n)$. The auxiliary random variable $U$ denotes the common message.

***Theorem 1:*** The capacity region of the two-user linear Gaussian compound MAC with common message and with ISI is given by

$$\mathcal{C}(P_1, P_2) =$$

$$\bigcup_{\substack{0 \leq \alpha_q(\omega) \leq 1 \\ \frac{1}{2\pi}\int_{-\pi}^{\pi} P_q(\omega)d\omega \leq P_q \\ q=1,2.}} \left\{ \begin{array}{l} (R_0, R_1, R_2): R_0 \geq 0, R_1 \geq 0, R_2 \geq 0, \\ R_1 \leq \min\{\mathbb{I}_1, \mathbb{I}_2\} \\ R_2 \leq \min\{\mathbb{I}_3, \mathbb{I}_4\} \\ R_1 + R_2 \leq \min\{\mathbb{I}_5, \mathbb{I}_6\} \\ R_0 + R_1 + R_2 \leq \min\{\mathbb{I}_7, \mathbb{I}_8\} \end{array} \right\}$$
(13)

where the terms $\mathbb{I}_i, i \in \{1, \dots, 8\}$ are defined as (14), shown at the top of the next page. The union is over power allocation across all the parallel sub-channels.

*Proof:* Refer to Section IV.

### B. Capacity Region of the Linear Gaussian Strong Interference Channel with Common Message and with ISI

We now consider the Gaussian strong interference channel with common message (SICC) and with ISI. Similar above, we can define linear Gaussian SICC (LGSICC) with finite memory $m$, and its equivalent $n$-block *memoryless circular* channel model called $n$-block circular Gaussian SICC (*n*-CGSICC*)*, for $n > m$. In fact, an $n$-CGCMACC is an $n$-CGSICC if

$$I(x_1^n; y_1^n | x_2^n, u^n) \leq I(x_1^n; y_2^n | x_2^n, u^n) \quad (15\text{-}1)$$
$$I(x_2^n; y_1^n | x_1^n, u^n) \leq I(x_2^n; y_2^n | x_1^n, u^n) \quad (15\text{-}2)$$

Therefore, we can easily derive the capacity region of the $n$-CGSICC using (11) and (15), and using it we can obtain the capacity region of the linear Gaussian SICC (LGSICC) with ISI.

$$\mathbb{I}_1 = \frac{1}{4\pi}\int_{-\pi}^{\pi} \log\left(1 + \frac{\bar{\alpha}_1(\omega)P_1(\omega)|\tilde{H}_{11}(\omega)|^2}{\tilde{N}_1(\omega)}\right)d\omega \tag{14-1}$$

$$\mathbb{I}_2 = \frac{1}{4\pi}\int_{-\pi}^{\pi} \log\left(1 + \frac{\bar{\alpha}_1(\omega)P_1(\omega)|\tilde{H}_{12}(\omega)|^2}{\tilde{N}_2(\omega)}\right)d\omega \tag{14-2}$$

$$\mathbb{I}_3 = \frac{1}{4\pi}\int_{-\pi}^{\pi} \log\left(1 + \frac{\bar{\alpha}_2(\omega)P_2(\omega)|\tilde{H}_{21}(\omega)|^2}{\tilde{N}_1(\omega)}\right)d\omega \tag{14-3}$$

$$\mathbb{I}_4 = \frac{1}{4\pi}\int_{-\pi}^{\pi} \log\left(1 + \frac{\bar{\alpha}_2(\omega)P_2(\omega)|\tilde{H}_{22}(\omega)|^2}{\tilde{N}_2(\omega)}\right)d\omega \tag{14-4}$$

$$\mathbb{I}_5 = \frac{1}{4\pi}\int_{-\pi}^{\pi} \log\left(1 + \frac{\bar{\alpha}_1(\omega)P_1(\omega)|\tilde{H}_{11}(\omega)|^2 + \bar{\alpha}_2(\omega)P_2(\omega)|\tilde{H}_{21}(\omega)|^2}{\tilde{N}_1(\omega)}\right)d\omega \tag{14-5}$$

$$\mathbb{I}_6 = \frac{1}{4\pi}\int_{-\pi}^{\pi} \log\left(1 + \frac{\bar{\alpha}_1(\omega)P_1(\omega)|\tilde{H}_{12}(\omega)|^2 + \bar{\alpha}_2(\omega)P_2(\omega)|\tilde{H}_{22}(\omega)|^2}{\tilde{N}_2(\omega)}\right)d\omega \tag{14-6}$$

$$\mathbb{I}_7 = \frac{1}{4\pi}\int_{-\pi}^{\pi} \log\left(1 + \frac{P_1(\omega)|\tilde{H}_{11}(\omega)|^2 + P_2(\omega)|\tilde{H}_{21}(\omega)|^2 + \sqrt{\alpha_1(\omega)\alpha_2(\omega)P_1(\omega)P_2(\omega)}\left(2\mathrm{Re}\{\tilde{H}_{11}(\omega)\tilde{H}_{21}^{\dagger}(\omega)\}\right)}{\tilde{N}_1(\omega)}\right)d\omega \tag{14-7}$$

$$\mathbb{I}_8 = \frac{1}{4\pi}\int_{-\pi}^{\pi} \log\left(1 + \frac{P_1(\omega)|\tilde{H}_{12}(\omega)|^2 + P_2(\omega)|\tilde{H}_{22}(\omega)|^2 + \sqrt{\alpha_1(\omega)\alpha_2(\omega)P_1(\omega)P_2(\omega)}\left(2\mathrm{Re}\{\tilde{H}_{12}(\omega)\tilde{H}_{22}^{\dagger}(\omega)\}\right)}{\tilde{N}_2(\omega)}\right)d\omega \tag{14-8}$$

---

**Corollary 1:** The capacity region of the two-user linear Gaussian strong interference channel with common message and ISI is given by

$$\mathcal{C}(P_1, P_2) = \bigcup_{\substack{0 \leq \alpha_q(\omega) \leq 1 \\ \frac{1}{2\pi}\int_{-\pi}^{\pi} P_q(\omega)d\omega \leq P_q \\ q=1,2.}} \left\{ \begin{array}{l} (R_0, R_1, R_2): R_0 \geq 0, R_1 \geq 0, R_2 \geq 0, \\ R_1 \leq \mathbb{I}_1 \\ R_2 \leq \mathbb{I}_4 \\ R_1 + R_2 \leq \min\{\mathbb{I}_5, \mathbb{I}_6\} \\ R_0 + R_1 + R_2 \leq \min\{\mathbb{I}_7, \mathbb{I}_8\} \end{array} \right\} \tag{16}$$

where the terms $\mathbb{I}_i, i \in \{1,4,5,6,7,8\}$ are defined as (14), shown at the top of the page.

### IV. PROOF OF THEOREM 1

Let $W_0^G$, $W_1^G$, and $W_2^G$ be random variables distributed according to $\mathcal{CN}(0,1)$. We first define the following mappings. The superscript $G$ is used to denote Gaussian distribution.

$$U_k^G = \left(\sqrt{\alpha_1 P_1} + \sqrt{\alpha_2 P_2}\right) W_0^G$$
$$X_{1,k}^G = \sqrt{\alpha_1 P_1} W_0^G + \sqrt{\bar{\alpha}_1 P_1} W_1^G$$
$$X_{2,k}^G = \sqrt{\alpha_2 P_2} W_0^G + \sqrt{\bar{\alpha}_2 P_2} W_2^G$$

where $\alpha_q \in [0,1]$ and $\bar{\alpha}_q = 1 - \alpha_q$ for $q \in \{1,2\}$. By using these mappings, and considering the channel model described by (8) and (9), we obtain

$$\tilde{Y}_{1,k}^G = \tilde{H}_{11,k} X_{1,k} + \tilde{H}_{21,k} X_{2,k} + \tilde{Z}_{1,k}$$

$$= \tilde{H}_{11,k}\left(\sqrt{\alpha_1 P_1} W_0^G + \sqrt{\bar{\alpha}_1 P_1} W_1^G\right)$$
$$+ \tilde{H}_{21,k}\left(\sqrt{\alpha_2 P_2} W_0^G + \sqrt{\bar{\alpha}_2 P_2} W_2^G\right) + \tilde{Z}_{1,k} \tag{17}$$

$$\tilde{Y}_{2,k}^G = \tilde{H}_{12,k} X_{1,k} + \tilde{H}_{22,k} X_{2,k} + \tilde{Z}_{2,k}$$
$$= \tilde{H}_{12,k}\left(\sqrt{\alpha_1 P_1} W_0^G + \sqrt{\bar{\alpha}_1 P_1} W_1^G\right)$$
$$+ \tilde{H}_{22,k}\left(\sqrt{\alpha_2 P_2} W_0^G + \sqrt{\bar{\alpha}_2 P_2} W_2^G\right) + \tilde{Z}_{2,k} \tag{18}$$

By the invertibility of the DFT and considering (17) and (18), we evaluate the mutual information terms in (11) as follows. As we know for any real sequence $d^n$, its DFT $D^n$ has the property that $D_k = D_{n-k}^*, 1 \leq k \leq n$, where $D^*$ denotes the complex conjugate of $D$. Thus, without losing any information, we can reconstruct the entire sequence $d^n$ using the DFT terms $\{D_1, \ldots, D_l\}$, where $l = \left\lfloor \frac{n}{2} \right\rfloor$. Therefore, we have:

$$I(x_1^n; y_1^n | x_2^n, u^n) = I(X_1^l; \tilde{Y}_1^l | X_2^l, U^l)$$
$$= \sum_{k=1}^{l} I(X_{1,k}; \tilde{Y}_{1,k} | X_{2,k}, U_k)$$
$$= \sum_{k=1}^{l} \{h(\tilde{Y}_{1,k} | X_{2,k}, U_k) - h(\tilde{Y}_{1,k} | X_{1,k}, X_{2,k}, U_k)\}$$
$$\leq \sum_{k=1}^{l} \{h(\tilde{Y}_{1,k}^G | X_{2,k}^G, U_k^G) - h(\tilde{Y}_{1,k} | X_{1,k}^G, X_{2,k}^G, U_k^G)\} \tag{19}$$
$$= \sum_{k=1}^{l} \frac{1}{2} \log\left(\frac{|\mathrm{cov}(\tilde{H}_{11,k}(\sqrt{\bar{\alpha}_1 P_1} W_1^G) + \tilde{Z}_{1,k})|}{|\mathrm{cov}(\tilde{Z}_{1,k})|}\right)$$

$$= \sum_{k=1}^{l} \frac{1}{2} \log\left(\frac{\bar{\alpha}_1(\omega_k)P_1(\omega_k)|\tilde{H}_{11}(\omega_k)|^2 + \tilde{N}_1(\omega_k)}{\tilde{N}_1(\omega_k)}\right)$$

$$= \sum_{k=1}^{l} \frac{1}{2} \log\left(1 + \frac{\bar{\alpha}_1(\omega_k)P_1(\omega_k)|\tilde{H}_{11}(\omega_k)|^2}{\tilde{N}_1(\omega_k)}\right)$$

where (19) follows from the fact that a Gaussian distribution maximizes the entropy. Similarly, we can evaluate other terms in (11). Therefore, $C_n(P_1, P_2)$ can be expressed as

$$C_n(P_1, P_2) = \bigcup_{\substack{0 \leq \alpha_q(\omega_k) \leq 1 \\ \frac{1}{n}\sum_1^n P_q(\omega_k) \leq P_q \\ q=1,2.}} \begin{Bmatrix} (R_0, R_1, R_2): R_0 \geq 0, R_1 \geq 0, R_2 \geq 0, \\ R_1 \leq \min\{T_1, T_2\} \\ R_2 \leq \min\{T_3, T_4\} \\ R_1 + R_2 \leq \min\{T_5, T_6\} \\ R_0 + R_1 + R_2 \leq \min\{T_7, T_8\} \end{Bmatrix}$$

(20)

where the terms $T_i, i \in \{1, ..., 8\}$ are defined as (21), shown at the bottom of the page. The $\tilde{H}_{pq}(\omega_k), p, q \in \{1,2\}$ is the $k$th-component channel for the link $pq$; the $\tilde{N}_q(\omega_k)$ is similarly defined noise components. The $P_q(\omega_k)$ is the total power allocated to the $k$th-component channel by the user $q$, and $\alpha_q(\omega_k)$ is the fraction of $P_q(\omega_k)$ allocated to the user $q$ on channel $k$ for common message, and $\bar{\alpha}_q(\omega_k)$ is the fraction of $P_q(\omega_k)$ allocated to the user $q$ on channel $k$ for private message. Finally, by using properties of Riemann integration, we reach to the desired result in the limit as $n \to \infty$.

---

$$T_1 = \sum_{k=0}^{l} \frac{1}{2n} \log\left(1 + \frac{\bar{\alpha}_1(\omega_k)P_1(\omega_k)|\tilde{H}_{11}(\omega_k)|^2}{\tilde{N}_1(\omega_k)}\right)$$

$$T_2 = \sum_{k=0}^{l} \frac{1}{2n} \log\left(1 + \frac{\bar{\alpha}_1(\omega_k)P_1(\omega_k)|\tilde{H}_{12}(\omega_k)|^2}{\tilde{N}_2(\omega_k)}\right)$$

$$T_3 = \sum_{k=0}^{l} \frac{1}{2n} \log\left(1 + \frac{\bar{\alpha}_2(\omega_k)P_2(\omega_k)|\tilde{H}_{21}(\omega_k)|^2}{\tilde{N}_1(\omega_k)}\right)$$

$$T_4 = \sum_{k=0}^{l} \frac{1}{2n} \log\left(1 + \frac{\bar{\alpha}_2(\omega_k)P_2(\omega_k)|\tilde{H}_{22}(\omega_k)|^2}{\tilde{N}_2(\omega_k)}\right)$$

$$T_5 = \sum_{k=0}^{l} \frac{1}{2n} \log\left(1 + \frac{\bar{\alpha}_1(\omega_k)P_1(\omega_k)|\tilde{H}_{11}(\omega_k)|^2 + \bar{\alpha}_2(\omega_k)P_2(\omega_k)|\tilde{H}_{21}(\omega_k)|^2}{\tilde{N}_1(\omega_k)}\right)$$

$$T_6 = \sum_{k=0}^{l} \frac{1}{2n} \log\left(1 + \frac{\bar{\alpha}_1(\omega_k)P_1(\omega_k)|\tilde{H}_{12}(\omega_k)|^2 + \bar{\alpha}_2(\omega_k)P_2(\omega_k)|\tilde{H}_{22}(\omega_k)|^2}{\tilde{N}_2(\omega_k)}\right)$$

$$T_7 = \sum_{k=0}^{l} \frac{1}{2n} \log\left(1 + \frac{P_1(\omega_k)|\tilde{H}_{11}(\omega_k)|^2 + P_2(\omega_k)|\tilde{H}_{21}(\omega_k)|^2 + \sqrt{\alpha_1(\omega_k)\alpha_2(\omega_k)P_1(\omega_k)P_2(\omega_k)}\left(2\text{Re}\{\tilde{H}_{11}(\omega_k)\tilde{H}_{21}^\dagger(\omega_k)\}\right)}{\tilde{N}_1(\omega_k)}\right)$$

$$T_8 = \sum_{k=0}^{l} \frac{1}{2n} \log\left(1 + \frac{P_1(\omega_k)|\tilde{H}_{12}(\omega_k)|^2 + P_2(\omega_k)|\tilde{H}_{22}(\omega_k)|^2 + \sqrt{\alpha_1(\omega_k)\alpha_2(\omega_k)P_1(\omega_k)P_2(\omega_k)}\left(2\text{Re}\{\tilde{H}_{12}(\omega_k)\tilde{H}_{22}^\dagger(\omega_k)\}\right)}{\tilde{N}_2(\omega_k)}\right)$$

(21)